\documentclass[twocolumn,letterpaper]{IEEEAerospaceCLS}  % only supports two-column, letterpaper format

\usepackage[]{graphicx}    % We use this package in this document
\newcommand{\ignore}[1]{}  % {} empty inside = %% comment

\usepackage{pgfplots}
\pgfplotsset{compat=1.17}
\usepgfplotslibrary{groupplots}
\usepackage{tikz}
\usetikzlibrary{external}
\usepackage{tikzscale}

\usepackage{amsmath, amssymb} % math package
\usepackage{tensor}
\usepackage{mathrsfs} % script L
\usepackage{graphicx}
\usepackage{caption}
\usepackage{hyperref}
\usepackage{color}
\usepackage{optidef}
\usepackage{siunitx}
\usepackage{cuted}
\usepackage{xcolor}
\usepackage[normalem]{ulem}
\usepackage[ruled, linesnumbered]{algorithm2e}

\usepackage[capitalise]{cleveref}
\crefname{equation}{}{} % this makes \cref{equation} match \eqref
\Crefname{equation}{Equation}{Equations}

\DeclareMathOperator{\sign}{sign}

\begin{document}

\title{Building a Better B-Dot: Fast Detumbling with Non-Monotonic Lyapunov Functions}

\author{%
Jacob B. Willis\\ 
Robotics Institute\\
Carnegie Mellon University\\
5000 Forbes Ave.\\
Pittsburgh, PA 15213\\
jbwillis@cmu.edu
\and 
Paulo R.M. Fisch\\ 
Robotics Institute\\
Carnegie Mellon University\\
5000 Forbes Ave.\\
Pittsburgh, PA 15213\\
pfisch@andrew.cmu.edu
\and 
Aleksei Seletskiy\\ 
Robotics Institute\\
Carnegie Mellon University\\
5000 Forbes Ave.\\
Pittsburgh, PA 15213\\
aseletsk@andrew.cmu.edu
\and 
Zachary Manchester\\
Robotics Institute\\
Carnegie Mellon University\\
5000 Forbes Ave.\\
Pittsburgh, PA 15213\\
zacm@cmu.edu
%%%% IMPORTANT: Use the correct copyright information--IEEE, Crown, or U.S. government. %%%%%
\thanks{\footnotesize 979-8-3503-0462-6/24/$\$31.00$ \copyright2024 IEEE}
}

\maketitle

\thispagestyle{plain}
\pagestyle{plain}

\maketitle

\thispagestyle{plain}
\pagestyle{plain}

\begin{abstract}
Spacecraft detumbling with magnetic torque coils is an inherently underactuated control problem.
Contemporary and classical magnetorquer detumbling methods do not adequately consider this underactuation, and suffer from poor performance as a result.
These controllers can get stuck on an uncontrollable manifold, resulting in long detumbling times and high power consumption.
This work presents a novel detumble controller based on a non-monotonic Lyapunov function that predicts the future magnetic field along the satellite's orbit and avoids uncontrollable configurations.
In comparison to other controllers in the literature, our controller detumbles a satellite in significantly less time while also converging to lower overall angular momentum.
We provide a derivation and proof of convergence for our controller as well as Monte-Carlo simulation results demonstrating its performance in representative use cases.
\end{abstract}

\tableofcontents

\section{Introduction}
After a spacecraft is deployed on orbit, a common first phase of operation is detumbling.
In this phase, the angular velocity of the satellite is reduced from tens of degrees per second to rates that are tolerated by the satellite mission or managed by other onboard control systems.
To perform detumbling, the spacecraft must reduce its total angular momentum by one to two orders of magnitude.
This is only accomplished by generating external torques, either through expending propellant, or, in low-Earth orbit, with magnetic torque coils (magnetorquers) that exchange momentum with the Earth’s magnetic field.

Magnetorquers are appealing because they do not require expending propellant.
However, magnetorquers are underactuated: at any instant in time, they only generate torque in a two-dimensional subspace perpendicular to the Earth’s local magnetic-field vector.
To prove convergence, most common magnetorquer detumbling controllers, including the classic B-dot and B-cross controllers~\cite{markley1978attitude,markley2014fundamentals}, rely on the motion of the satellite through the Earth’s magnetic field, which makes the magnetic field vector time varying in the orbit frame. 
While instantaneously underactuated, over the spacecraft's full orbit, full controllability is achieved.
In this work, we demonstrate that these classic controllers, and their modern variants, can take many hours to detumble a spacecraft, despite it being possible to detumble much faster and with much less total control effort.
For this reason, these controllers are inefficient, wasting precious energy and time during the critical early stages of satellite operation.

To mitigate the inefficiencies of the classic magnetic detumbling controllers, we present a novel controller that uses a prediction of the future magnetic field vector, dramatically improving convergence time.
The magnetic field prediction is done using only gyroscope and magnetometer sensor measurements; no inertial attitude or position reference is required.
The controller is based on a discrete-time non-monotonic Lyapunov function~\cite{ahmadi2008non}, which is able to temporarily increase the angular momentum of the spacecraft, allowing the system to move away from control singularities.
We compare our controller against controllers in the literature using a Monte-Carlo simulation of 100 randomly sampled initial conditions. 
Our controller detumbles a satellite in less time than other controllers; it also converges to lower overall angular rates.

Our contributions include:
\vspace{-\baselineskip}
\begin{itemize}
    \item A unified treatment of the numerous magnetorquer detumbling controllers that exist in the literature
    \item A derivation and analytic proof of convergence of our predictive detumbling controller based on a discrete-time non-monotonic Lyapunov function
    \item Monte-Carlo simulation experiments showing the performance of our predictive controller in comparison to five other controllers in the literature
\end{itemize}

The paper proceeds as follows:
In \cref{sec:related_work} we discuss prior work on magnetorquer detumbling. 
In \cref{sec:background} we present the attitude dynamics of a spacecraft and provide a unified derivation of the five detumbling controllers that we compare ours to.
In \Cref{sec:control} we provide a brief introduction to non-monotonic Lyapunov functions and derive our discrete-time non-monotonic detumbling controller.
We present our Monte-Carlo simulation results in \cref{sec:simulation}, and we summarize our conclusions and suggest directions for future work in \cref{sec:conclusion}.

\section{Related Work} \label{sec:related_work}
Magnetorquer detumbling has a long history dating back to the earliest days of space exploration~\cite{markley1978attitude,stickler1976elementary}.
In general, magnetorquer detumbling controllers come in two categories with many variants: B-dot and B-cross.
B-dot controllers assume only magnetometer measurements are available. B-cross~\cite{markley2014fundamentals,avanzini2012magnetic} controllers assume both magnetometer and gyroscope measurements are available onboard the spacecraft.
As we will show in \cref{sec:bg_detumbling}, these two categories are related by a simple approximation, and the many variations in the literature reduce to a variety of gains and saturation methods for handling control limits~\cite{invernizzi2020projection,desouky2020new,desouky2020time}.

In addition to the magnetic detumbling methods discussed here, there has been significant work on full magnetic attitude control, including the work by Wisniewski~\cite{wisniewski1999optimal} which models the magnetic field as a periodic system, and more recent work that utilizes numerical optimal control to perform three-axis magnetorquer attitude control~\cite{silani2005magnetic,gatherer2019magnetorquer}.
Ovchinnikov presents a recent survey of both magnetorquer detumbling and attitude control~\cite{ovchinnikov2019survey}.

\section{Background} \label{sec:background}

\subsection{Attitude Dynamics}
Let $h \in {\mathbb{R}}^3$ be the angular momentum of a spacecraft, $B \in {\mathbb{R}}^3$ be the Earth's local geomagnetic field vector at the spacecraft's location, and $\mu \in {\mathbb{R}}^3$ be the dipole moment produced by the spacecraft's magnetorquers.

With a magnetic dipole moment as input, a spacecraft's angular momentum dynamics expressed in an inertial reference frame are
\begin{align} \label{eq:hdot}
    \dot{h} = \tau = -B \times \mu = -\hat{B}\mu
\end{align}
where $\tau \in {\mathbb{R}}^3$ is the torque on the spacecraft and $\hat{B}$ is the skew-symmetric cross product matrix,
\begin{align}
    \hat{B} &= \begin{bmatrix}
        0 & -B_z & B_y \\
        B_z & 0 & -B_x \\
        -B_y & B_x & 0
    \end{bmatrix}.
\end{align}

With inertia matrix $J \in {\mathbb{R}}^{3 \times 3}$, the inertial angular momentum relates to the inertial angular velocity $\omega$ as
\begin{align}
    h = J \omega.
\end{align}

We make use of the time derivative of the geomagnetic field vector with respect to the spacecraft's body frame, $\dot{B}^{\mathcal{B}}$, and with respect to the inertial frame, $\dot{B}^{\mathcal{N}}$. Both are expressed in body-fixed coordinates.
The relationship between these quantities is
\begin{align}
    \dot{B}^{\mathcal{N}} &= \hat{\omega} B + \dot{B}^{\mathcal{B}}.\label{eq:BdotN}
\end{align}

\subsection{Detumbling Control}\label{sec:bg_detumbling}
Many of the detumbling control laws found in the literature are variations of a single control law that is derived from the Lyapunov function
\begin{align}
    V = \frac12 h^Th \label{eq:continuous_momentum_lyapunov}.
\end{align}
Taking the time derivative,
\begin{align}
\dot{V} = h^T \dot{h} = -h^T\hat{B}\mu.
\end{align}
We desire to find $\mu$ that minimizes $\dot{V}$ at every instant in time.
To do so, we formulate this as an optimization problem with bound constraints that limit the maximum dipole moment the satellite can produce:
\begin{mini}<b> 
  {\mu}{\dot{V} = -h^T\hat{B}\mu}{}{}
  \addConstraint{-\mu_{\mathrm{max}}}{\leq \bar{\mu} \leq } { \mu_{\mathrm{max}}}. \labelOP{opt:control_lyapunov_continuous}
\end{mini} 
% \begin{equation}\label{opt:control_lyapunov_continuous}
% \begin{aligned}
%     &\min_\mu &&  \dot{V} = -h^T\hat{B}\mu \\
%     &{\mathrm{s.t.}} && -\mu_{\mathrm{max}} \leq \bar{\mu} \leq \mu_{\mathrm{max}}.
% \end{aligned} 
% \end{equation}
This optimization problem is a linear program with a closed-form solution in the form of a bang-bang control law:
\begin{align}
    \mu =  \mu_{\mathrm{max}}\sign(\hat{h} B) ,\label{eq:time_optimal}
\end{align}
where the $\sign$ function is interpreted element-wise.
Bang-bang controllers like \cref{eq:time_optimal} are prone to chattering in the presence of noise, so we replace the $\sign$ hard saturation with a soft saturation,
\begin{align}
    \mu =  \mu_{\mathrm{max}}\tanh(k \hat{h} B), \label{eq:lyapunov_momentum}
\end{align}
where the $\tanh$ function is, again, interpreted element-wise and $k$ is a tuning parameter. We refer to this control law as the Lyapunov momentum control law.

The control law in \cref{eq:time_optimal} is closely related to the classical B-dot and B-cross control laws~\cite{markley2014fundamentals}.
The B-cross law replaces $h$ with $\omega$ and relaxes the bang-bang saturation to a linear feedback law with gain $k$,
\begin{align}
    \mu = k \hat{\omega} B. \label{eq:bcross}
\end{align}
Avanzini and Giulietti~\cite{avanzini2012magnetic} propose selecting the B-cross controller gain, 
\begin{align}
    k = 2 \frac{1}{\sqrt{a^3 / GM}} (1 + \sin(\xi_m)) \lambda_{\mathrm{min}}, \label{eq:bcross_gain}
\end{align}
where $a$ is the orbit semi-major axis, $GM$ is the Earth's gravitational parameter, $\xi_m$ is the orbit's geomagnetic inclination and $\lambda_{\mathrm{min}}$ is the minimum eigenvalue of the spacecraft's inertia matrix $J$.

The B-dot law~\cite{markley1978attitude,stickler1976elementary} modifies \cref{eq:bcross} by making the assumption that $\dot{B}^{\mathcal{N}} = 0$ so 
\begin{align}
    \hat{\omega}B \approx -\dot{B}^{\mathcal{B}}, \label{eq:B_approx_omega_cross_B}
\end{align}
resulting in 
\begin{align}
    \mu = -k {\dot{B}}^{\mathcal{B}}. \label{eq:bdot}
\end{align}
The B-dot law has the advantage that $\dot{B}^{\mathcal{B}}$ is readily estimated from a magnetometer only, so no gyroscope measurements are required for its implementation.
However, as $\omega \to 0$, the approximation in \cref{eq:B_approx_omega_cross_B} becomes less accurate and the B-dot law tends to converge to larger final momentum.

Desouky~\cite{desouky2020new,desouky2020time} presented two control laws: 
Their ``time-optimal'' control law is equivalent to \cref{eq:time_optimal} and their ``B-dot Variant'' control law inverts \cref{eq:B_approx_omega_cross_B} with a regularizing term to solve for $\omega$ and substitutes the result into \cref{eq:bcross} to obtain the control law,
\begin{align}
    \mu = -k \hat{B} (\epsilon I + \hat{B})^{-1} \dot{B}^{\mathcal{B}}, \label{eq:bdot_variant}
\end{align}
where $0 < \epsilon \ll 1$, and $\epsilon = 1\times 10^{-6}$ in practice.

Invernizzi and Lovera~\cite{invernizzi2020projection} use a projection-based method to compute a time-varying gain for an unsaturated version of \cref{eq:time_optimal},
\begin{equation}\label{eq:projection_detumble}
\begin{aligned}
    \mu &= - \frac{k}{\|B\|^2} \hat{B} h,\\
    k &= k_1\exp \left(-k_2 \left| \dfrac{B^T h}{\|B\| (\|h\| + \epsilon)}\right|\right).
\end{aligned}
\end{equation}

All of the previously discussed methods are closely related and suffer from the same fundamental limitation: when the controlled variable ($h$ or $\omega$) and $B$ are aligned their cross product is zero and the commanded control input goes to zero.
The controllers are convergent on long time scales because $B$ is time varying in the orbital frame, so eventually the cross product will no longer be zero.
However, they are prone to tracking this uncontrollable subspace, effectively becoming stuck and taking significantly longer to detumble.

\begin{figure}[!htb]
    \centering
    \includegraphics[width=0.9\columnwidth]{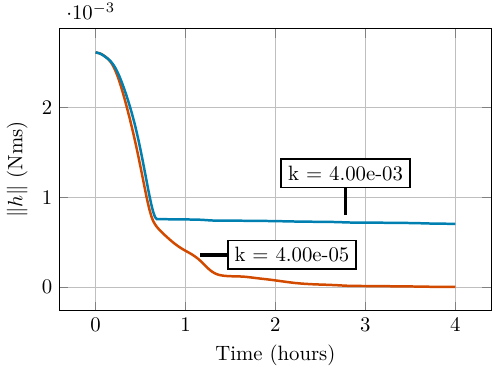}
    \caption{Two simulation runs of the B-cross controller in \cref{eq:bcross} with different gains. As the gain increases, the controller convergence gets worse because it gets stuck in the uncontrollable subspace where $\omega$ and $B$ are parallel.}
    \label{fig:bcross_stuck}
\end{figure}

Consider the B-cross controller in \cref{eq:bcross}. 
If we decompose $\omega = \omega^\parallel + \omega^\perp$, where $\omega^\parallel$ and $\omega^\perp$ are the components of $\omega$ parallel and perpendicular to $B$,
\begin{align}
    \mu = k(\omega^\parallel + \omega^\perp) \times B = k \hat{\omega}^\perp B ,
\end{align}
and
\begin{align}
    \dot{h} = -\hat{B} \mu = -k \hat{B} (\hat{\omega}^\perp B) = - \lambda \omega^\perp ,
\end{align}
for some $\lambda > 0$.
Therefore, $h$ is only reduced in the $\omega^\perp$ direction with no change in the $\omega^\parallel$ direction.
\Cref{fig:bcross_stuck} shows this effect in two simulation runs of the B-cross controller with the same initial conditions and two different gains. 
The smaller gain was chosen based on \cref{eq:bcross_gain} and the larger gain is a factor of 100 larger.
At the beginning, the smaller-gain controller decreases $\|h\|$ at a slower rate, but ultimately converges sooner because the larger gain causes the controller to get stuck on the uncontrollable subspace where $\omega$ and $B$ are parallel.
The gain sweep results in \cref{fig:gain_sweep} show a similar phenomena occurring with the other controllers.
While this phenomenon can be partially mitigated with appropriate tuning, there are no guarantees that the controller will converge without becoming stuck.
The result is a longer-than-necessary convergence time and higher-than-necessary energy expenditure.

To address the shortcomings of existing controllers, we relax one of their basic constraints: we derive a controller that does not decrease the angular momentum of the spacecraft monotonically, but still maintains a Lyapunov convergence guarantee \emph{on average}.
Intuitively, this allows the controller to trade off increasing the angular momentum instantaneously in exchange for avoiding the uncontrollable subspace, making the angular momentum more controllable in the future.

\section{Non-Monotonic Controller Derivation} \label{sec:control}

We begin by introducing discrete-time monotonic Lyapunov analysis, then extend it to non-monotonic Lyapunov analysis.
The discrete-time dynamical system
\begin{align} \label{eq:discrete_dynamical}
    x_{k+1} = f(x_k)
\end{align}
with $x \in {\mathbb{R}}^n$ has a globally asymptotically stable (GAS) equilibrium at $x = 0$ if there exists a Lyapunov function $V(x): {\mathbb{R}}^n \to {\mathbb{R}}$ such that
\begin{subequations} \label{eq:lyapunov}
\begin{align}
    &V(x) > 0 &&\forall x \neq 0 \\
    &V(0) = 0 \\
    &V_{k+1} < V_k &&\forall k \label{eq:lyapunov_monotonic_decrease}
\end{align}
\end{subequations}
where we use the notation $V_k = V(x_k)$.
It is well known that there is no general method of finding a Lyapunov function that satisfies \cref{eq:lyapunov}, even if the system is GAS. 
Ahmadi and Parrilo suggest that the monotonic decrease condition in \cref{eq:lyapunov_monotonic_decrease} may be too restrictive, and present several alternative stability theorems that only require $V$ to decrease on average~\cite{ahmadi2008non}. 
We rely on Theorem 2.1 from their work.
It modifies the conditions in \cref{eq:lyapunov} so that \cref{eq:discrete_dynamical} is GAS at $x = 0$ if there exists a scalar $\alpha \geq 0$ and a Lyapunov function $V:{\mathbb{R}}^n \to {\mathbb{R}}$ such that
\begin{subequations} \label{eq:nmlyapunov}
\begin{align}
    &V(x) > 0 &&\forall x \neq 0 \label{eq:lyapunov_nonmonotonic_posdef}\\
    &V(0) = 0 \label{eq:lyapunov_nonmonotonic_zero}\\
    &\alpha (V_{k+2} - V_k) + (V_{k+1} - V_k) < 0 &&\forall k. \label{eq:lyapunov_nonmonotonic_decrease}
\end{align}
\end{subequations}
The condition in \cref{eq:lyapunov_nonmonotonic_decrease} relaxes \cref{eq:lyapunov_monotonic_decrease} to allow $V_k$ to decrease on average between two timesteps.

\subsection{Non-Monotonic Detumbling}
To derive the non-monotonic detumbling controller, we begin with the discrete-time Lyapunov function:
\begin{align}
    V_k = \frac{1}{2} h_k^T h_k.
\end{align}
This trivially satisfies \cref{eq:lyapunov_nonmonotonic_posdef,eq:lyapunov_nonmonotonic_zero}, so it remains to design the control input $\mu$ such that the non-monotonic Lyapunov condition 
\begin{align}
    \Delta V = \alpha (V_{k+2} - V_k) + (V_{k+1} - V_k) < 0
\end{align}
from \cref{eq:lyapunov_nonmonotonic_decrease} is satisfied for $\alpha \geq 0$.

After discretizing the attitude dynamics in time with Euler integration and expanding as shown in the \hyperref[sec:DV_expansion]{Appendix}, we find that
\begin{align}
\Delta V = \frac12\bar{\mu}^T (Q_1 + \alpha Q_2) \bar{\mu} - (q_1 + \alpha q_2)^T\bar{\mu}
\end{align}
where $Q_1, Q_2 \in {\mathbb{R}}^{6\times 6}$ are symmetric positive semi-definite matrices, $\bar{\mu} \in {\mathbb{R}}^6 = [\mu_k^T, \mu_{k+1}^T]^T$ is the vector of control inputs at $k$ and $k+1$, and $q_1, q_2 \in {\mathbb{R}}^6$.
This means that $\Delta V$ is convex, and, as we will see in the following, its minimum is less than zero.
So, it is possible to find $\bar{\mu}$ such that $\Delta V < 0$, and the non-monotonic Lyapunov conditions of \cref{eq:lyapunov_nonmonotonic_decrease} are satisfied.

We wish to find a control law for $\bar{\mu}$ such that $\Delta V$ is minimized:
\begin{mini}<b> 
  {\bar{\mu}}
  {\Delta V = \frac12\bar{\mu}^T (Q_1 + \alpha Q_2) \bar{\mu} - (q_1 + \alpha q_2)^T\bar{\mu}}
  {}{}
  \addConstraint{-\mu_{\mathrm{max}}}{\leq \bar{\mu} \leq } { \mu_{\mathrm{max}}} \labelOP{opt:minDV}.
\end{mini} 
% \begin{equation} \label{opt:minDV}
% \begin{aligned}
%   &\min_\bar{\mu} &&\Delta V = \frac12\bar{\mu}^T (Q_1 + \alpha Q_2) \bar{\mu} - (q_1 + \alpha q_2)^T\bar{\mu} \\
%   &\mathrm{s.t.} && -\mu_\mathrm{max} \leq \bar{\mu} \leq \mu_\mathrm{max}.
% \end{aligned}
% \end{equation} 
Since $Q_1,$ and $Q_2$ are positive semi-definite, $\Delta V$ is not strictly convex and \cref{opt:minDV} has multiple minima. 
We add $\frac{1}{2} \beta \bar{\mu}^T \bar{\mu} $ with $1 \gg \beta > 0$ as a regularizing term to make the objective strictly convex. 
The result is a convex quadratic program that is reliably and quickly solved with a numerical solver.
Alternatively, we can make the same simplification as in \cref{eq:lyapunov_momentum} and solve the unconstrained minimization problem, enforcing a soft saturation constraint on the result.
The optimization is then
\begin{mini}<b> 
  {\bar{\mu}}
  {\begin{aligned}F = &\frac12 \beta \bar{\mu}^T \bar{\mu} + \frac12\bar{\mu}^T (Q_1 + \alpha Q_2) \bar{\mu} \\&- (q_1 + \alpha q_2)^T\bar{\mu}\end{aligned}}
  {}{} \labelOP{opt:minF}.
\end{mini} 
% \begin{equation}\label{opt:minF}
% \begin{aligned}
%   &\min_\bar{\mu} &&F = \frac12 \beta \bar{\mu}^T \bar{\mu} + \frac12\bar{\mu}^T (Q_1 + \alpha Q_2) \bar{\mu} - (q_1 + \alpha q_2)^T\bar{\mu}\\
%   &\mathrm{s.t.} &&-\mu_\mathrm{max} \leq \bar{\mu} \leq \mu_\mathrm{max}. 
% \end{aligned}
% \end{equation}
We find the analytic solution by taking the gradient of $F$ with respect to $\bar{\mu}$ and setting it to zero.
The gradient of $F$ is
\begin{align}
    \nabla F = \beta\bar{\mu} + (Q_1 + \alpha Q_2) \bar{\mu} - (q_1 + \alpha q_2). 
\end{align}
Setting to zero and solving for $\bar{\mu}$ gives our control law,
\begin{align}
   \bar{\mu}^* =  (\beta I + Q_1 + \alpha Q_2)^{-1} (q_1 + \alpha q_2). \label{eq:regularized_mbar}
\end{align}
% With only the identity matrix \cref{eq:regularized_mbar} is the momentum B-cross control law.
We now show that this control law satisfies the nonmonotonic Lyapunov decrease condition.
Plugging $\bar{\mu}^*$ into $F$ and recalling that $(I+Q_1 + Q_2)$ is symmetric, we have
\begin{align}
    % \begin{split}
    % F^* &= \frac12 (q_1 + \alpha q_2)^T (\beta I + Q_1 + \alpha Q_2)^{-1} (\beta I + Q_1 + \alpha Q_2) (\beta I + Q_1 + \alpha Q_2)^{-1} (q_1 + \alpha q_2)\\
    % &\quad- (q_1 + \alpha q_2)^T (\beta I + Q_1 + \alpha Q_2)^{-1} (q_1 + \alpha Q_2) \\
    % \end{split}\\
    % \begin{split}
    % &= \frac12 (q_1 + \alpha q_2)^T (\beta I + Q_1 + \alpha Q_2)^{-1} (q_1 + \alpha q_2)\\
    % &\quad- (q_1 + \alpha q_2)^T (\beta I + Q_1 + \alpha Q_2)^{-1} (q_1 + \alpha q_2) \\
    % \end{split}\\
    F^* &= -\frac12 (q_1 + \alpha q_2)^T (\beta I + Q_1 + \alpha Q_2)^{-1} (q_1 + \alpha q_2).
\end{align}
Since $\beta I + Q_1 + \alpha Q_2$ is positive definite, $(\beta I + Q_1 + \alpha Q_2)^{-1}$ is also positive definite, and $F^* < 0$ for all $\alpha, \beta > 0$.
The regularizing term in $F$ is always positive in $\bar{\mu}$, so we can conclude that $\Delta V < 0$, which satisfies the nonmonotonic Lyapunov decrease condition in~\cref{eq:lyapunov_nonmonotonic_decrease}.

% \jbw{I think there is another condition here that $B_k$ and $B_{k+1}$ are not parallel?} The solution still holds

\subsection{Causal Implementation}\label{sec:causal_control}
Examining \cref{eq:B_bar_def}, we see that the $Q_1$ and $Q_2$ matrices rely on knowledge of $B_{k+1}$. 
This is not causal. 
However, $B_{k+1}$ can be predicted using knowledge of the satellite's orbit and a model of the geomagnetic field.
Detumbling is often executed during early operations of a satellite, so orbit knowledge and a computationally expensive geomagnetic field model may not be available.
An alternative is to approximate $B_{k+1}$ as
\begin{align}
    B_{k+1} \approx B_{k} + \Delta t \dot{B}_k^{\mathcal{N}}. \label{eq:B_2}
\end{align}
We cannot directly measure $\dot{B}^{\mathcal{N}}$. However, using \cref{eq:BdotN}, it can be estimated from multiple magnetometer and gyro measurements.
%To do so, a high-quality estimate of $\dot{B}^{\mathcal{B}}$ can be made using multiple measurements of $B$.
%For ease of implementation, we currently use a derivative of the magnetic field model computed using automatic differentiation to compute $\dot{B}^{\mathcal{B}}$.

\subsection{Complete Controller}
Bringing together the development from the last two sections, the discrete non-monotonic controller is given in \cref{alg:discrete_nonmonotonic}.
The input $B_{k+1}$ is computed using the approximation in \cref{eq:B_2}.
On \cref{alg:normalize_b1,alg:normalize_b2}, we normalize the values of $B_*$ to avoid numerical issues and ensure consistency of performance across the wide range of geomagnetic field magnitudes a spacecraft will experience. 
\Crefrange{alg:bar_b}{alg:q_2} set up the problem components and follow from the derivation in the \hyperref[sec:DV_expansion]{Appendix}.
We compute $\bar{\mu}$ on \cref{alg:mu_bar_solve} by solving a linear system.
Finally, on \cref{alg:tanh_saturation} we perform a soft saturation of the computed control output and rescale it to satisfy the satellite's control limits.

\begin{algorithm}
\caption{Discrete Non-monotonic controller}
\label{alg:discrete_nonmonotonic}
\KwData{Given $B_k$, $B_{k+1}$, $J$, $\omega$, $k$, $\alpha$, $\beta$}
\KwResult{$\mu$}
$\qquad b_1 = B_k / \|B_k\|$ \label{alg:normalize_b1}\\
$\qquad b_2 = B_{k+1} / \|B_{k+1}\|$ \label{alg:normalize_b2}\\
$\qquad \bar{b} = \begin{bmatrix} \hat{b_1} & \hat{b_2} \end{bmatrix}$\label{alg:bar_b}\\
$\qquad \footnotesize Z = \beta\begin{bmatrix} I & 0 \\ 0 & 0 \end{bmatrix}$\\
$\qquad Q_1 = Z \bar{b} \bar{b}^T Z$\\
$\qquad Q_2 = \alpha \bar{b} \bar{b}^T$\\
$\qquad h = J \omega$\\
$\qquad q_1 = Z \bar{b} h$\\
$\qquad q_2 = \alpha \bar{b} h$ \label{alg:q_2}\\
$\qquad \bar{\mu} = (I + Q_1 + Q_2) \backslash (q_1 + q_2)$\label{alg:mu_bar_solve}\\
$\qquad \mu = \mu_{\mathrm{max}}\tanh(k \bar{\mu}[1\!:\!3])$ \label{alg:tanh_saturation}\\
\end{algorithm}

% \subsection{Differentiating Magnetometer Measurements}\label{sec:magnetometer_differentiation}
% A high-quality estimate of ${}^V B$ and ${}^V \dot{B}$ is essential for the performance of the controller. To compute this estimate, we use a sliding-window linear fit of the magnetometer measurements.
% For a set of vehicle-frame measurements ${}^V \mathcal{B}_{k-4}, \dots, {}^V \mathcal{B}_k$ taken at uniform times $h = t_k - t_{k-1}$, the filtered measurement is
% \begin{align}
%     {}^V B = \frac{6{}^V \mathcal{B}_k + 4{}^V \mathcal{B}_{k-1} + 2{}^V \mathcal{B}_{k-2} - 2 {}^V \mathcal{B}_{k-4}}{10} \label{eq:measurement_filter}
% \end{align}
% and the derivative estimate is
% \begin{align}
%     {}^V \dot{B} = \frac{2{}^V \mathcal{B}_k + {}^V \mathcal{B}_{k-1} - {}^V \mathcal{B}_{k-3} - 2 {}^V \mathcal{B}_{k-4}}{10h} \label{eq:derivative_estimate}.
% \end{align}

\section{Simulation Experiments} \label{sec:simulation}
\setcounter{footnote}{0}

\begin{figure*}[!t]
    \centering
    \includegraphics[width=\linewidth]{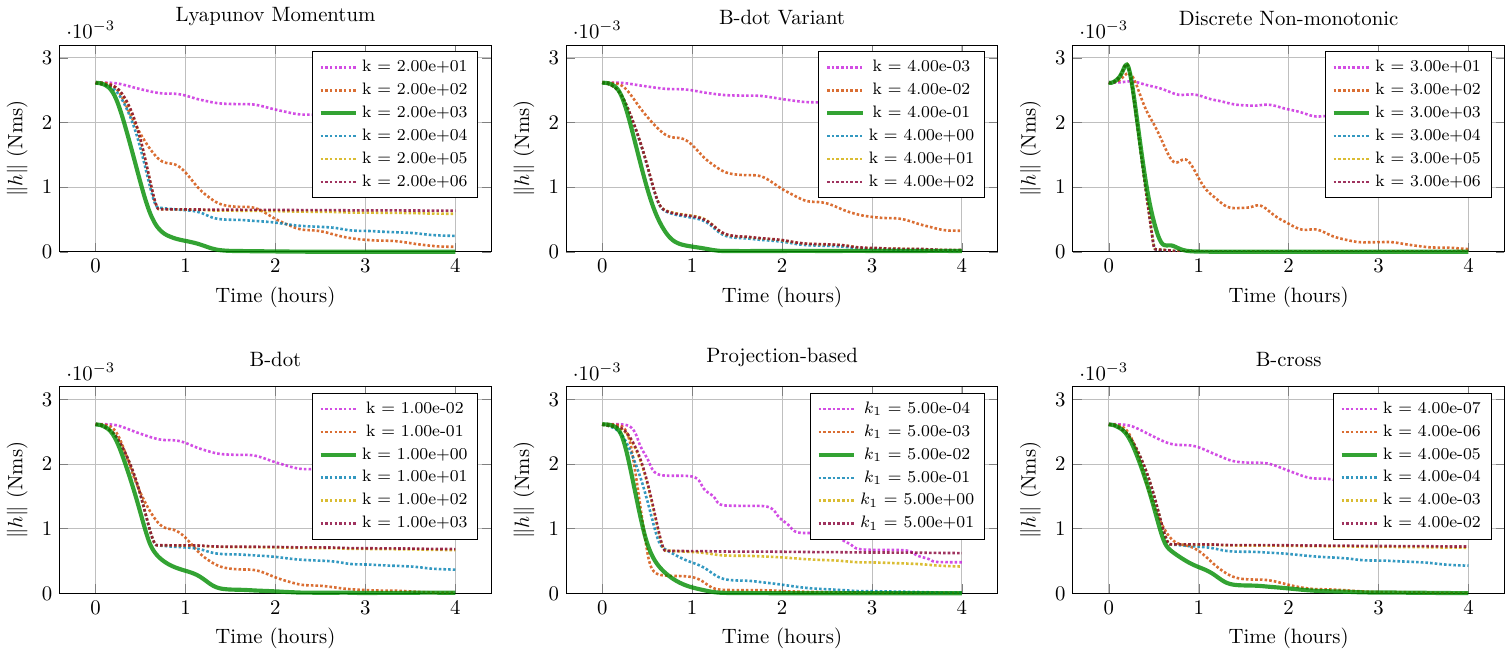}
    \caption{Gain sweep study showing the effect each controller's gain has on its detumbling performance for a single initial condition. The solid green line is the gain that was used for the Monte-Carlo simulation experiment shown in \cref{fig:detumble_time_histogram,fig:momentum_vs_time,fig:final_magnitude_histogram}. The other lines are for gains varying from two orders of magnitude lower to three orders of magnitude higher than the chosen gain.}
    \label{fig:gain_sweep}
\end{figure*}

\begin{figure*}[!t]
    \centering
    \includegraphics[width=\linewidth]{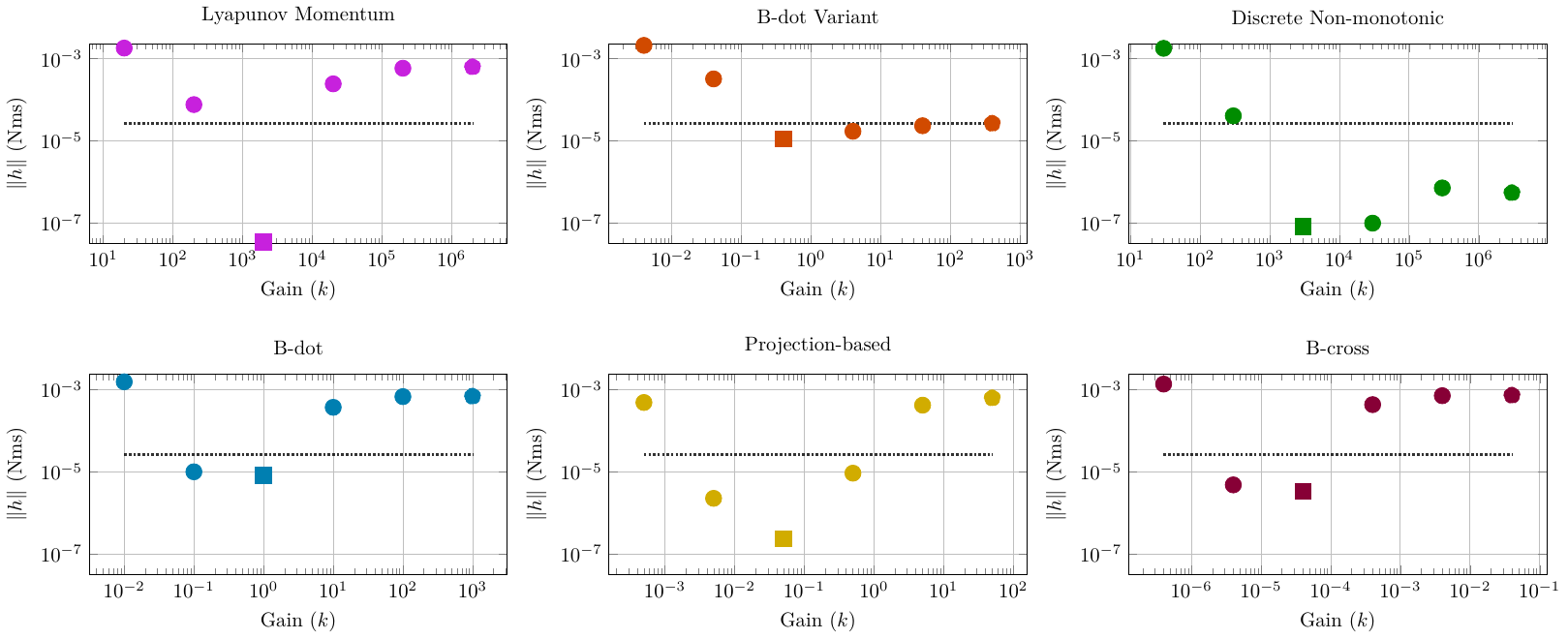}
    \caption{Final momentum magnitude for the gain sweep trajectories shown in \cref{fig:gain_sweep}. 
     The dotted horizontal line indicates the 1\% threshold used to determine convergence.
     The square marker corresponds to the gain used in the Monte-Carlo simulation experiment shown in \cref{fig:detumble_time_histogram,fig:momentum_vs_time,fig:final_magnitude_histogram}.
    }
    \label{fig:gain_sweep_final_h}
\end{figure*}

\begin{figure*}[!t]
    \centering
    \includegraphics[width=\linewidth]{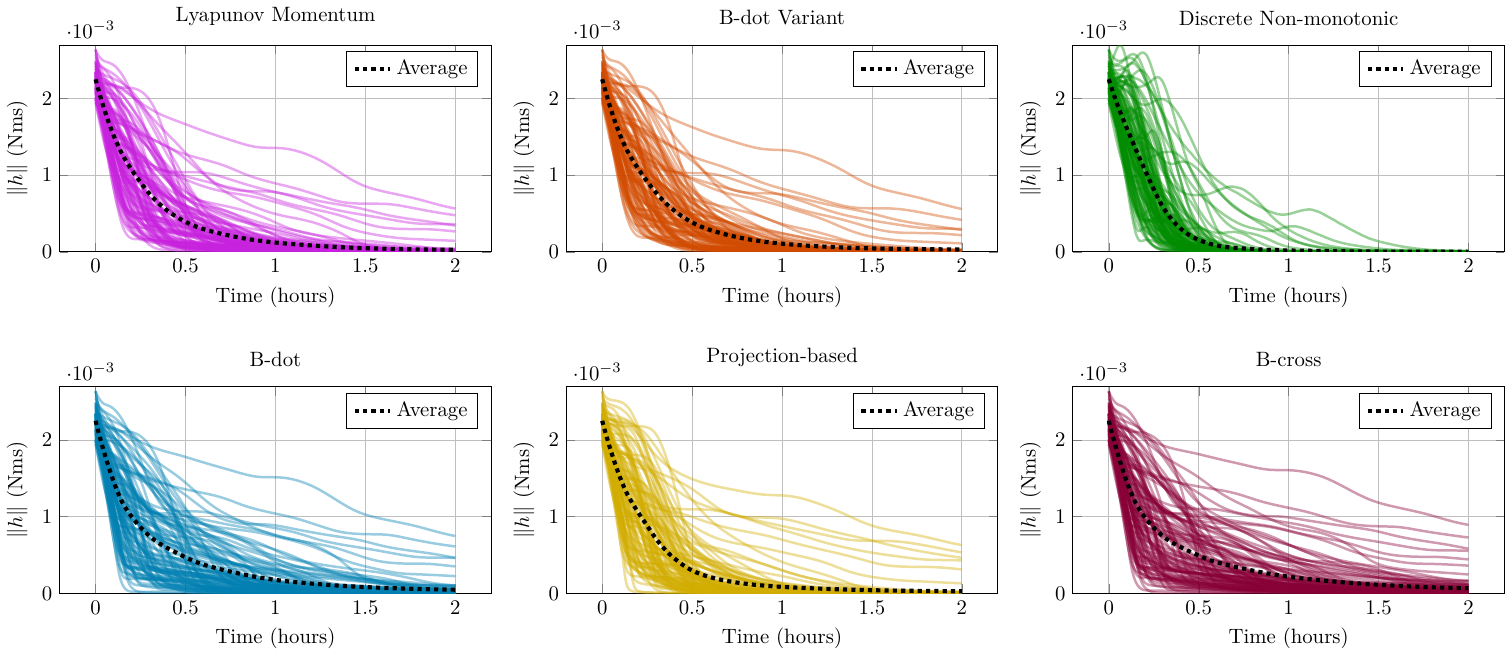}
    \caption{Momentum magnitude versus time plot for each of the controllers discussed in this paper. The Discrete Non-monotonic controller differs from the other controllers in that the system momentum increases before decreasing and converging to zero. This is the key distinction of this controller and allows it to have faster convergence times than other detumbling controllers.}
    \label{fig:momentum_vs_time}
\end{figure*}

\begin{figure*}[!t]
    \centering
    \includegraphics[width=\linewidth]{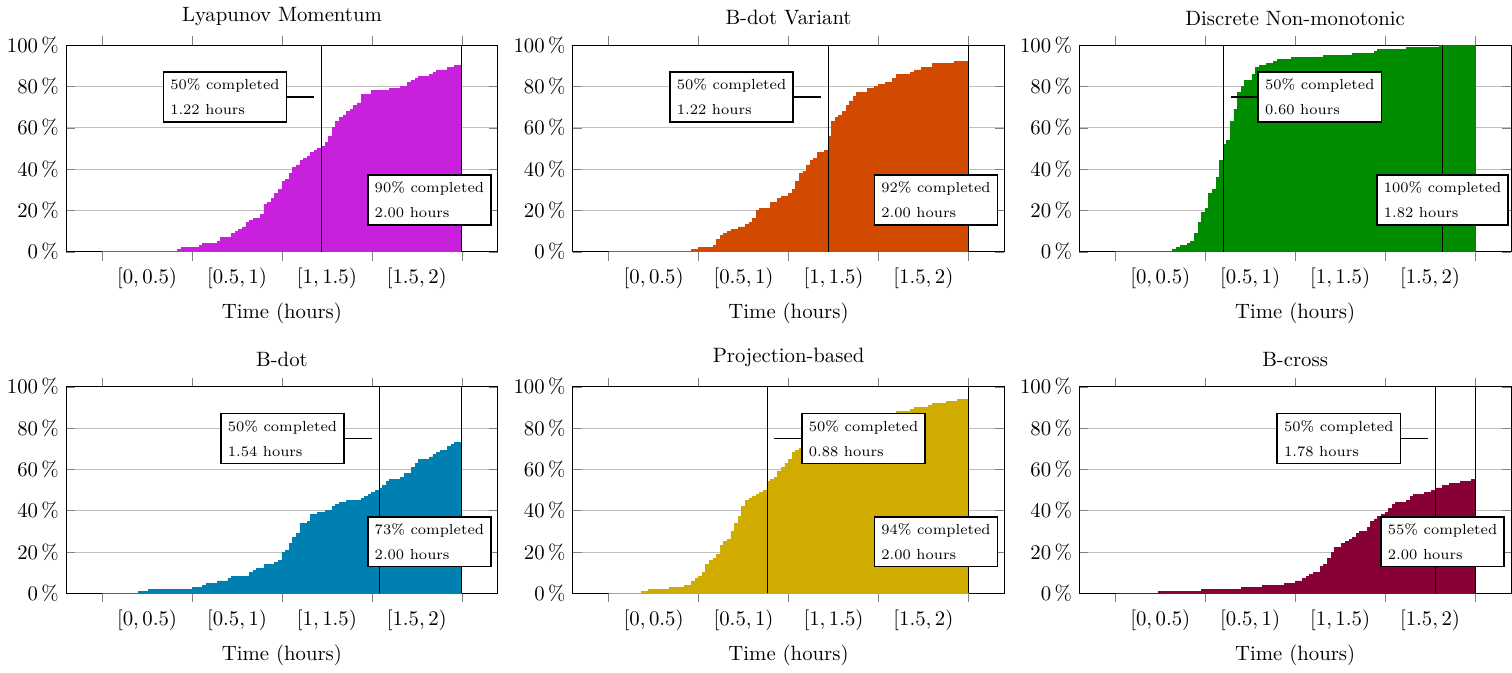}
    \caption{Cumulative distribution of detumble times for each of the controllers discussed in this paper. Detumble times are defined as the time when the satellite first reaches 1\% of its initial angular momentum. Each simulation run ended at two hours, so only detumble times less than two hours are counted. Our Discrete Non-monotonic controller has the lowest average detumble time; it is also the only controller to detumble for all simulation runs.}
    \label{fig:detumble_time_histogram}
\end{figure*}

\begin{figure*}[!t]
    \centering
    \includegraphics[width=\linewidth]{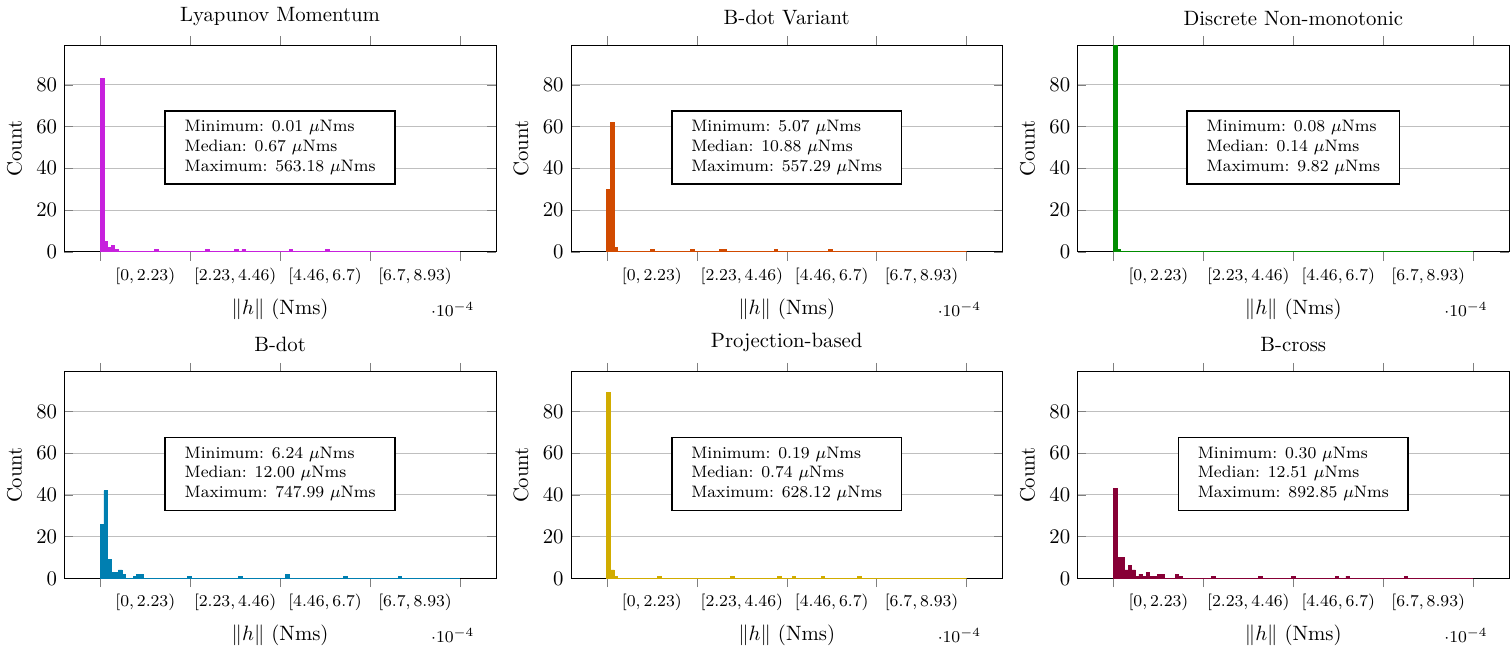}
    \caption{Histogram of momentum magnitudes after 2 hours for each of the controllers discussed in this paper. Our Discrete Non-monotonic controller has lower median and maximum momentum magnitude than any of the other controllers.}
    \label{fig:final_magnitude_histogram}
\end{figure*}
All simulations are performed in a 12-degree-of-freedom orbital-and-attitude-dynamics simulation. 
All code is available on GitHub\footnote{ \href{https://github.com/RoboticExplorationLab/non-monotonic-detumbling/}{github.com/RoboticExplorationLab/non-monotonic-detumbling}}.
The simulation environment relies on the open-source SatelliteDynamics.jl\footnote{\href{https://sisl.github.io/SatelliteDynamics.jl/latest/}{sisl.github.io/SatelliteDynamics.jl}} orbital dynamics package and includes perturbations due to J2 and atmospheric drag. 
The attitude dynamics include orbit-coupled drag torques.
To accurately model the geomagnetic field, we use the International Geomagnetic Reference Field (IGRF)~\cite{alken2021international}.
The spacecraft properties used for the simulations are given in \cref{tab:spacecraft_properties}; they reflect the properties for a 1.5U CubeSat with printed circuit board magnetorquers embedded in the solar panels.
Noisy sensor measurements and a randomly initialized constant gyro bias are also included in the simulation; the noise parameters are representative of low-cost micro-electromechanical (MEMS) gyro and magnetometer hardware.

\subsection{Gain Sweep Study}
Each of the controllers is sensitive to its tuning parameter $k$.
To provide meaningful comparisons between controllers, each controller needs to be tuned to perform in the best possible manner.
To do so, we simulate the performance impact of each controller's gain, sweeping it over several orders of magnitude.
The results of this study are shown in \cref{fig:gain_sweep,fig:gain_sweep_final_h}.
The solid green line in \cref{fig:gain_sweep} is the gain that was used for the Monte-Carlo simulation experiment; this gain was chosen as a tradeoff between fast convergence and avoiding high gains that lead to the controller getting stuck in their uncontrollable subspace.

\Cref{fig:gain_sweep_final_h} shows the final momentum of each of the trajectories from \cref{fig:gain_sweep}, plotted against their corresponding gains.
This allows us to more clearly see the performance of the controllers when under- and over-tuned.
The Lyapunov momentum, B-dot, projection-based, and B-cross controllers fail to converge when the gain is too low or too high.
The convergence failure with high gains is another example of the controllers getting stuck on their uncontrollable manifold as discussed in \cref{sec:bg_detumbling}.
This suggests that tuning these controllers to converge consistently could be challenging.
In contrast, the B-dot variant  shows consistent convergence once the gain is at or above $4.00\times 10^{-1}$ but it converges to the highest final momentum of all of the controllers when well tuned.
Our discrete non-monotonic controller remains well under the 1\% threshold for all investigated gains, suggesting it is more robust to poor tuning than the other controllers.

\subsection{Monte-Carlo Simulation Experiments}
For each controller, the Monte-Carlo simulation starts from each of 100 randomly sampled initial states.
Initializing each controller with the same random initial states allows for fair comparison.
The ranges and values of the Monte-Carlo initial conditions are sampled from a uniform distribution with minimum and maximum values given by the ranges in \cref{tab:mc_initial_conditions}.
They represent random circular orbits at a fixed altitude and random vehicle axis of rotation with a fixed initial angular velocity magnitude.
To avoid the lack of controllability all magnetorquer control systems experience at near equatorial inclinations~\cite{bhat2005controllability}, we restricted the orbital inclinations to $[20\si{\degree}, 160\si{\degree})$. 
Since most satellites in low-Earth orbit operate at high inclinations we believe these results translate well to real orbital configurations.
The controller parameters and corresponding equation reference are shown in \cref{tab:controller_parameters}.

The simulation results are shown in \cref{fig:detumble_time_histogram,fig:momentum_vs_time,fig:final_magnitude_histogram}.
\Cref{fig:detumble_time_histogram} shows a histogram of the time it takes for the spaceraft momentum to be reduced to 1\% of its initial value.
Many of the simulation runs of the common control methods do not converge to this 1\% threshold within two hours.
However, our discrete non-monotonic controller converges within the two hour simulation period for all initial conditions tested, with the majority of the initial conditions converging in one hour or less.
The reason for this can be seen in \cref{fig:momentum_vs_time}; it shows the time history of the momentum magnitude for the 100 Monte-Carlo simulation runs, as well as an average time history of these runs.
The discrete non-monotonic controller exhibits significantly different behavior than the five other controllers, increasing in momentum one or more times before finally converging to zero.

\Cref{fig:final_magnitude_histogram} shows the final angular momentum magnitudes.
The discrete non-monotonic controller has the smallest maximum final angular momentum and the smallest median final angular momentum.
Since the controllers were all stopped at two hours regardless of convergence, the median provides a better point of comparison than the maximum.
The discrete non-monotonic, Lyapunov momenum and projection-based controllers have median momentum less than $0.1 \mu \mathrm{Nms}$ while the B-dot, B-dot variant, and B-cross controllers all result in a median final angular momentum that is more than 10 times larger.
The B-dot and B-dot variant controllers also have a minimum final angular momentum that is more than 10 times the other controllers.
This suggests that angular velocity information from a gyroscope is useful in achieving a smaller final angular momentum.

\begin{table}[htb]
    \centering
    \caption{Simulated spacecraft properties}
    \begin{tabular}{l|l}
        Name & Value \\
         \hline\hline
         Inertia $J_{xx}$ & \num{4.5e-3} \si{\kilogram\meter\squared} \\
         $J_{xy}$ & -\num{3.2e-4} \si{\kilogram\meter\squared}\\
         $J_{xz}$ &  0.0 \si{\kilogram\meter\squared}\\
         $J_{yy}$ & \num{5.1e-3} \si{\kilogram\meter\squared} \\
         $J_{yz}$ & 0.0 \si{\kilogram\meter\squared} \\
         $J_{zz}$ & \num{3.7e-3} \si{\kilogram\meter\squared} \\
         Mass & 1.6 \si{\kilogram} \\
         Dimensions & 75 \si{\centi\meter} $\times$ 10 \si{\centi\meter} $\times$ 15 \si{\centi\meter} \\
         Drag Coefficient & 2.2 \\
         $\mu_{\mathrm{max}}$ &  $\begin{bmatrix} 0.070 & 0.053 & 0.070 \end{bmatrix}$ \si{\ampere\meter\squared}\\
         Magnetometer noise & 15 \si{\nano\tesla} \\
         Gyro noise & $0.005$\si{\degree\per\second\per\sqrt{\hertz}}\\
         Gyro bias & $1$ \si{\degree\per\second}
    \end{tabular}
    \label{tab:spacecraft_properties}
\end{table}

\begin{table}[htb]
    \centering
    \caption{Monte-Carlo Initial Condition Distribution}
    \begin{tabular}{l|l}
        Satellite State & Value \\
         \hline\hline
         Altitude & 400 \si{\kilo\meter} \\
         Eccentricity & 0 \\
         Inclination & $[20\si{\degree}, 160\si{\degree})$ \\
         RAAN & $[0\si{\degree}, 360\si{\degree})$ \\
         Arg. of Latitude & $[0\si{\degree}, 360\si{\degree})$ \\
         $\|\omega \|$ & $30\si{\degree\per\second}$
    \end{tabular}
    \label{tab:mc_initial_conditions}
\end{table}

\begin{table}[htb]
    \centering
    \caption{Controller Parameters}
    \begin{tabular}{l|l}
         Controller & Parameters \\
         \hline\hline
         Lyapunonv Momentum~\cref{eq:lyapunov_momentum} & $k = \num{2.0e3}$\\
         B-cross~\cref{eq:bcross} & $k = \num{4.0e-6}$\\
         Discrete Non-monotonic & $k=\num{3.0e3}$, $\alpha = 100$,\\
         \multicolumn{1}{r|}{(\cref{alg:discrete_nonmonotonic})}& $\beta = 1$, $\Delta t = 10\;\si{\minute}$\\
         B-dot Variant~\cref{eq:bdot_variant} & $k = \num{0.4}$, $\epsilon = \num{1e-6}$\\
         Projection-based~\cref{eq:projection_detumble} & $k_1 = \num{0.05},$ $k_2 = \num{4.0}$,\\
         & $\epsilon = \num{1e-8}$\\
         B-dot~\cref{eq:bdot} & $k = 1.0$
    \end{tabular}
    \label{tab:controller_parameters}
\end{table}

% \subsection{Worst-Case Initial Conditions}
% \jbw{After doing a lot of simulations I'm not sure this is actually the worst case...}
% Because of the cross product in \cref{eq:hdot}, it is not possible to produce a torque that is aligned with $B$.
% For this reason, the worst-case initial state for a satellite's momentum is for it to be along the maximum principle inertial axis and for that axis to be aligned with $B$.
% Let $b = B / \|B\|$ in the inertial frame, and $\nu$ in the body frame be the maximum principle axis of the satellite, that is, the Eigenvector of $J$ with the largest corresponding Eigenvalue.
% Then we choose the initial body to inertial attitude to align $\nu$ with $b$ by finding the axis-angle rotation from $\nu$ to $b$.
% The rotation axis is
% \begin{align}
% \phi = \frac{\nu \times b}{\|\nu \times b\|}
% \end{align}
% and the rotation angle is
% \begin{align}
%     \theta = \cos^{-1}(\nu^T b)
% \end{align}
% resulting in the body to inertial quaternion
% \begin{align}
%     q_0 = \begin{bmatrix}
%         \sin\left(\frac{\theta}{2}\right) \\ \cos\left(\frac{\theta}{2}\right) \phi
%     \end{bmatrix}
% \end{align}
% and the body-frame angular velocity
% \begin{align}
%     \omega_0 = \bar{\omega}_0 J^{-1} \nu
% \end{align}
% with $\bar{\omega}_0$ the initial angular velocity magnitude.

\section{Conclusions} \label{sec:conclusion}
The many variants of B-dot and B-cross controllers in the literature differ primarily in how the controller gains and saturation are selected. 
Their performance is similar, with each having the potential to get stuck in the uncontrollable subspace where the controlled state (angular momentum or angular velocity) aligns with the magnetic field vector.
Recent magnetorquer detumbling controllers, such as the projection-based controller~\cite{invernizzi2020projection}, improve on this failure mode but still exhibit similar worst-case convergence.
The novel non-monotonic Lyapunov magnetorquer detumbling control law we have presented is a more significant departure from the classical B-dot and B-cross control laws:  our control law implicitly predicts the future controllability of the system and avoids putting the satellite in an uncontrollable state.
In our Monte-Carlo simulations, it achieves detumbling times that are more than twice as fast as the other controllers while operating with realistic sensor noise and gyro bias.
In addition, our control law is straightforward to tune and less sensitive to tuning than other control laws.

To put our novel control law into practical use, a high-quality estimate of the time-derivative of the geomagnetic field is needed.
Future work will focus on generating this estimate and analyzing the full closed-loop performance of the magnetic field estimator and control law in combination.

% \appendix
% \section{Expansion of Non-monotonic Lyapunov Condition} \label{sec:DV_expansion}
\appendix{} \label{sec:DV_expansion}
From \cref{sec:control}, we have
\begin{align}
    \Delta V = \alpha (V_{k+2} - V_k) + (V_{k+1} - V_k) < 0 ,
\end{align}
and
\begin{align}
    V_k = \frac{1}{2} h_k^T h_k.
\end{align}
Through the rest of this section we drop the subscript $k$ and use $[\cdot]_0 = [\cdot]_k, [\cdot]_1 = [\cdot]_{k+1}, [\cdot]_2 = [\cdot]_{k+2}$ for clarity.

We approximate the discrete time dynamics in \cref{eq:hdot} using Euler integration, so that
\begin{subequations}
\begin{align}
    h_{1} &\approx h_0 + \Delta t \dot{h}_0 \\
    &= h_0 + \Delta t \tau_0 \\ 
    &= h_0 + \Delta t (\mu_0 \times B_0)
\end{align}
\end{subequations}
and
\begin{subequations}
\begin{align}
    h_{2} &\approx h_{1} + \Delta t \dot{h}_{1} \\
    &= h_0 + \Delta t \tau_0 + \Delta t \tau_{1} \\
    &= h_0 + \Delta t (\mu_0 \times B_0) + \Delta t (\mu_{1} \times B_{1}).
\end{align}
\end{subequations}
Substituting,
\begin{subequations}
\begin{align}
    V_{1} &= \frac12 h_{1}^T h_{1} \\
    % &= \frac12 \left (h_k + \Delta t \tau_k\right)^T \left(h_k + \Delta t \tau_k \right)\\
    % &= \frac12 \big(h_k^T h_k + \Delta t h_k^T\tau_k + \Delta t \tau_k^T h_k + \Delta t^2 \tau_k^T \tau_k \big) \\
    &= \frac12 \left( h_0^T h_0 + 2\Delta t h_0^T\tau_0 + \Delta t^2 \tau_0^T \tau_0 \right)
\end{align}
\end{subequations}
and
\begin{subequations}
\begin{align}
    V_{2} &= \frac12 h_{2}^T h_{2} \\
    % &= \frac12 \left (h_k + \Delta t \tau_k + \Delta t \tau_{k+1} \right)^T \left(h_k + \Delta t \tau_k + \Delta t \tau_{k+1} \right)\\
    % \begin{split}
    % &= \frac12 \big(h_k^T h_k + \Delta t h_k^T\tau_k + \Delta t h_k^T\tau_{k+1} \\ 
    % &\quad + \Delta t \tau_k^T h_k + \Delta t^2 \tau_k^T \tau_k + \Delta t^2 \tau_k^T \tau_{k+1} \\ 
    % &\quad + \Delta t \tau_{k+1}^T h_k + \Delta t^2 \tau_{k+1}^T \tau_k + \Delta t^2 \tau_{k+1}^T \tau_{k+1} \big)
    % \end{split}\\
    \begin{split}
    &= \frac12 \big( h_0^T h_0 + 2\Delta t h_0^T\tau_0 + 2\Delta t h_0^T\tau_{1} + \Delta t^2 \tau_0^T \tau_0 \\
    &\quad + 2\Delta t^2 \tau_0^T \tau_{1} + \Delta t^2 \tau_{1}^T \tau_{1} \big)
    \end{split}
\end{align}
\end{subequations}
so
\begin{subequations}
\begin{align}
\begin{split}
    V_{1} - V_0 &= \frac12 \left( h_0^T h_0 + 2\Delta t h_0^T\tau_0 + \Delta t^2 \tau_0^T \tau_0 \right)\\
    &\quad- \frac12 h_0^T h_0
\end{split}\\
    &= \Delta t h_0^T\tau_0 + \frac12 \Delta t^2 \tau_0^T \tau_0 \\
    &= -\Delta t h_0^T\hat{B}_0 \mu_0 + \frac12 \Delta t^2 \mu_0^T \hat{B}_0^T \hat{B}_0 \mu_0
\end{align}
\end{subequations}
and
\begin{subequations}
\begin{align}
    \begin{split}
       V_{2} - V_0 &= \frac12 \big( h_0^T h_0 + 2\Delta t h_0^T\tau_0 + 2\Delta t h_0^T\tau_{1} \\
       &\quad + \Delta t^2 \tau_0^T \tau_0 + 2\Delta t^2 \tau_0^T \tau_{1} + \Delta t^2 \tau_{1}^T \tau_{1} \big)\\
       &\quad - \frac12 h_0^T h_0
    \end{split}\\
    \begin{split}
        &= \Delta t h_0^T\tau_0 + \Delta t h_0^T\tau_{1} + \frac12 \Delta t^2 \tau_0^T \tau_0 \\
        &\quad+ \Delta t^2 \tau_0^T \tau_{1} + \frac12 \Delta t^2 \tau_{1}^T \tau_{1}
    \end{split}\\
    \begin{split}
    &= -\Delta t h_0^T \hat{B}_0 \mu_0 \\
    &\quad - \Delta t h_0^T \hat{B}_{1} \mu_{1} \\
    &\quad + \frac12 \Delta t^2 \mu_0^T \hat{B}_0^T \hat{B}_0 \mu_0 \\
    &\quad + \Delta t^2 \mu_0^T \hat{B}_0^T \hat{B}_{1} \mu_{1} \\
    &\quad + \frac12 \Delta t^2 \mu_{1}^T \hat{B}_{1}^T \hat{B}_{1} \mu_{1}
    \end{split}\\
\end{align}
\end{subequations}
where we used the identities
\begin{subequations}
\begin{align}
    \tau &= \mu \times B = - B \times \mu = - \hat{B} \mu\\
    \begin{split}
    \tau^T \tau &= (\mu \times B)^T (\mu \times B) = (- B \times \mu)^T (-B \times \mu) \\
    &= (- \hat{B} \mu)^T (- \hat{B} \mu) = \mu^T \hat{B}^T \hat{B} \mu.
    \end{split}
\end{align}
\end{subequations}
Bringing these terms together, we have
\begin{subequations}
\begin{align}
    \Delta V &\triangleq \alpha (V_{2} - V_0) + (V_{1} - V_0) \\
    \begin{split}
    &=\alpha \big(-\Delta t h_0^T \hat{B}_0 \mu_0 \\
    &\quad - \Delta t h_0^T \hat{B}_{1} \mu_{1} \\
    &\quad + \frac12 \Delta t^2 \mu_0^T \hat{B}_0^T \hat{B}_0 \mu_0 \\
    &\quad + \Delta t^2 \mu_0^T \hat{B}_0^T \hat{B}_{1} \mu_{1} \\
    &\quad + \frac12 \Delta t^2 \mu_{1}^T \hat{B}_{1}^T \hat{B}_{1} \mu_{1} \big)\\
    &\quad -\Delta t h_0^T\hat{B}_0 \mu_0 \\
    &\quad + \frac12 \Delta t^2 \mu_0^T \hat{B}_0^T \hat{B}_0 \mu_0
    \end{split}\\
    % \begin{split}
    % &= -(\alpha + 1)\Delta t h_0^T \hat{B}_0 \mu_0 \\
    % &\quad - \alpha \Delta t h_0^T \hat{B}_{1} \mu_{1} \\
    % &\quad + (\alpha + 1)\Delta t^2 \mu_0^T \hat{B}_0^T \hat{B}_0 \mu_0 \\
    % &\quad + \alpha \Delta t^2 \mu_0^T \hat{B}_0^T \hat{B}_{1} \mu_{1} \\
    % &\quad + \alpha \frac12 \Delta t^2 \mu_{1}^T \hat{B}_{1}^T \hat{B}_{1} \mu_{1}\\
    % \end{split}\\
    % \begin{split}
    % &= \frac12 \alpha \Delta t^2 \begin{bmatrix} \mu_0^T & \mu_{1}^T \end{bmatrix} 
    % \begin{bmatrix} \hat{B}_0^T \hat{B}_0 & \hat{B}_0^T \hat{B}_{1} \\ \hat{B}_{1}^T \hat{B}_0 & \hat{B}_{1}^T \hat{B}_{1} \end{bmatrix}
    % \begin{bmatrix} \mu_0 \\ \mu_{1} \end{bmatrix} \\ 
    % &\quad - \Delta t h_0^T \begin{bmatrix} \hat{B}_0^T & \hat{B}_{1}^T \end{bmatrix} 
    % \begin{bmatrix} \mu_0 \\ \mu_{1} \end{bmatrix} \\
    % %
    % &\quad + \frac12 \Delta t^2 \begin{bmatrix} \mu_0^T & \mu_{1}^T \end{bmatrix} 
    % \begin{bmatrix} \hat{B}_0^T \hat{B}_0 & 0 \\ 0 & 0 \end{bmatrix}
    % \begin{bmatrix} \mu_0 \\ \mu_{1} \end{bmatrix}\\
    % %
    % &\quad - \Delta t h_0^T \begin{bmatrix} \hat{B}_0 & 0 \end{bmatrix} 
    % \begin{bmatrix} \mu_0 \\ \mu_{1} \end{bmatrix}
    % \end{split}\\
    \begin{split}
    &= \alpha \frac12 \Delta t^2 \bar{\mu}^T \bar{B} \bar{B}^T \bar{\mu} \\
    &\quad - \alpha \Delta t h_0^T \bar{B}^T \bar{\mu} \\
    &\quad + \frac12 \Delta t^2 \bar{\mu}^T Z \bar{B} \bar{B}^T Z \bar{\mu} \\
    &\quad - \Delta t h_0^T \bar{B}^T Z \bar{\mu}
    \end{split}\\
    \begin{split}
    &= \frac12\bar{\mu}^T Q_1 \bar{\mu}  + \frac12 \alpha\bar{\mu}^T Q_2 \bar{\mu} \\
    &\quad - q_1^T\bar{\mu} - \alpha q_2^T\bar{\mu}
    \end{split}\\
    &= \frac12\bar{\mu}^T (Q_1 + \alpha Q_2) \bar{\mu} - (q_1 + \alpha q_2)^T\bar{\mu}
\end{align}
\end{subequations}
where we defined
\begin{subequations}\label{eq:nonmonotonic_ctrl_components}
\begin{align}
    \bar{\mu} &= \begin{bmatrix} \mu_0 \\ \mu_{1} \end{bmatrix} \in \mathbb{R}^6, \\
    \bar{B} &= \begin{bmatrix} \hat{B}_0^T \\ \hat{B}_{1}^T \end{bmatrix} \in \mathbb{R}^{6\times3} \label{eq:B_bar_def} \\
    Z &= \begin{bmatrix} I & 0 \\ 0 & 0 \end{bmatrix} \in \mathbb{R}^{3 \times 3} \\
    Q_1 &= \Delta t^2 Z \bar{B} \bar{B}^T Z\\
    Q_2 &= \alpha \Delta t^2 \bar{B} \bar{B}^T\\
    q_1 &= \Delta t (h_0^T \bar{B}^T Z)^T \\
    q_2 &= \alpha \Delta t (h_0^T \bar{B}^T)^T.
\end{align}
\end{subequations}
Since $\bar{B} \bar{B}^T$ is symmetric and ${\mathrm{rank}}(\bar{B} \bar{B}^T) = {\mathrm{rank}}(\bar{B}) \leq 3 < 6$, $Q_1$ and $Q_2$ are symmetric and positive semi-definite. 
%%%%%%%%%%%%%%%%%%%%%%%%%%%%%%%%%%%%%%%%%%%%%%%%%%%%%%%%%%%%%%%%%%%%%%%%%%%%%%%%%%%%%%%%%%%%%%%%%%%%%%
\acknowledgments
This work was supported by the Department of Defense National Defense Science and Engineering Graduate Fellowship (NDSEG) and by NASA under agreement 80NSSC21K0446.

The authors would also like to thank Davide Invernizzi for his correspondence and feedback. It significantly improved this paper.

\bibliographystyle{IEEEtran}
\bibliography{refs.bib}

%%%%%%%%%%%%%%%%%%%%%%%%%%%%%%%%%%%%%%%%%%%%%%%%%%%%%%%%%%%%%%%%%%%%%%%%%%%%%%%%%%%%%%%%%%%%%%%%%%%%%%
\vfill\null
% \columnbreak
\thebiography
%% This biostyle allows you to insert your photo size 1in X 1.25in
\begin{biographywithpic}
{Jacob Willis}{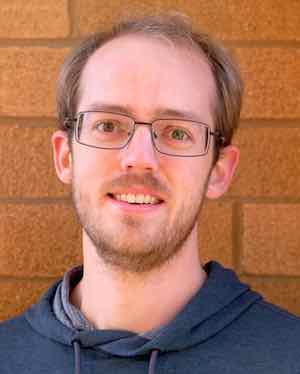}
is a PhD candidate in the Robotics Institute at Carnegie Mellon University. He is an NDSEG Fellow and received a BS and MS in Electrical and Computer Engineering from Brigham Young University in 2019 and 2021. His research interests include applications of numerical optimal control for autonomous aerospace systems, with recent work on attitude and formation control of small satellites.
\end{biographywithpic} 

\begin{biographywithpic}
{Paulo Fisch}{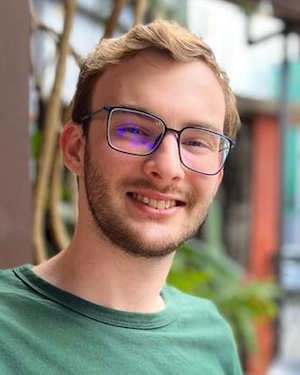}
is a PhD candidate in the Robotics Institute at Carnegie Mellon University. He has previous experience working at the German Aerospace Center (DLR) and got his Mechanical Engineering degree from the University of São Paulo in 2020. His interests include optimal state estimation for space systems and optimal control, with recent work on satellite orbit determination.
\end{biographywithpic} 

\begin{biographywithpic}
{Aleksei Seletskiy}{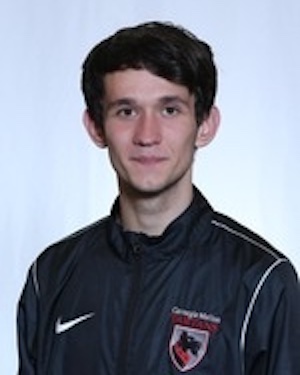}
is a Junior in Computer Science at Carnegie Mellon Uiversity.
His research interests include flight software, state estimation, and optimal control for satellite systems.
\vspace{0.5in}
\end{biographywithpic} 

\begin{biographywithpic}
{Zachary Manchester}{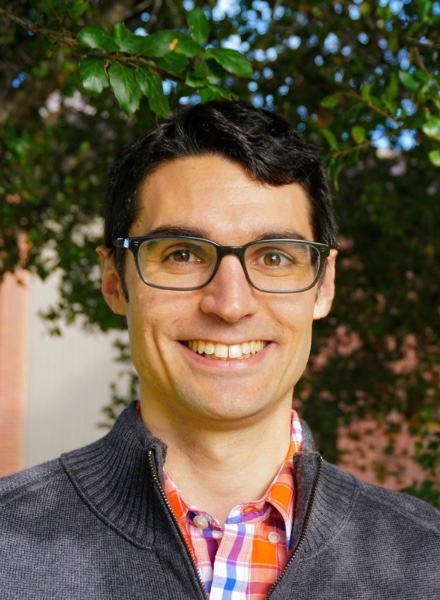}
is an assistant professor in the Robotics Institute at Carnegie Mellon University and founder of the Robotic Exploration Lab. He received a PhD in aerospace engineering in 2015 and a BS in applied physics in 2009, both from Cornell University. His research interests include control and optimization with application to aerospace and robotic systems with challenging nonlinear dynamics.
\end{biographywithpic}

\end{document}